\documentclass[a4paper,12pt]{article}

\usepackage{ifpdf}

\newif\ifpdf
\ifx\pdfoutput\undefined
  \pdffalse
\else
  \pdfoutput=1
  \pdftrue
\fi

\RequirePackage{xspace} %
\RequirePackage{subfigure} %
\RequirePackage[centertags]{amsmath} %
\RequirePackage{amssymb}
\RequirePackage{wrapfig} %
\RequirePackage{calc} %
\RequirePackage{ifthen}
\RequirePackage{tabularx} %
\RequirePackage{flafter} %
\RequirePackage{fancyhdr} %

\ifpdf
  \RequirePackage[pdftex]{color}%
  \RequirePackage{colortbl}%
  \RequirePackage{array}%
  \RequirePackage[pdftex]{graphicx}

  \RequirePackage[ pdftex, plainpages = false, pdfpagelabels,
                 pdfpagelayout = useoutlines,
                 bookmarks,
                 breaklinks = true,
                 linktocpage,
                 pagebackref,                      % to include page numbers in bibliography
                 colorlinks = true,
                 linkcolor = blue,
                 urlcolor  = blue,
                 citecolor = blue,
                 anchorcolor = blue,
                 hyperindex = true,
                 hyperfigures
                 ]{hyperref}
\else
  \RequirePackage{color}
  \RequirePackage{colortbl}
   \RequirePackage{array}
  \RequirePackage[dvips]{graphicx}
  \RequirePackage{hyperref}
  \usepackage{rotating}
\fi
\usepackage{makeidx} % enable indexing
\usepackage{setspace} %enable setting spacing, %e.g. singlespace, onehalfspace, and doublespace
\usepackage{rotating} %enable sidewaysfigure
\usepackage{ecltree}
\usepackage{epic}
\usepackage{supertabular}  % this will allow the table not to break
\usepackage{color}
\usepackage{exscale}
\usepackage{fontenc}
\usepackage{ifthen}
\usepackage{latexsym}
\usepackage{makeidx}
\usepackage{syntonly}
\usepackage{inputenc}
\usepackage{graphicx}
\usepackage{setspace}
\usepackage{caption2}
\usepackage[english]{babel}
\usepackage[square, comma,numbers,sort&compress]{natbib}
\usepackage{hypernat}
\usepackage{boxedminipage}
\usepackage{framed}
\usepackage{longtable}
\usepackage[all]{hypcap}

\newcommand{\note}[1]{\marginpar[left]{\singlespace \tiny #1}}

\renewcommand{\sectionmark}[1]%
      {\markright{\thesection\ #1}} %stops it capitalizing. #1 has value of section name

\renewcommand{\note}[1]{}
\doublespace % \onehalfspace %

\title
{ %
\vspace*{2.5cm} \LARGE{\bf Using Euler-Lagrange Variational Principle to Obtain
Flow Relations for Generalized Newtonian Fluids} \vspace*{3.5cm} \\
}

\author{Taha Sochi\footnote{University College London - Department of Physics \& Astronomy - Gower Street - London - WC1E 6BT.
Email: t.sochi@ucl.ac.uk.} \vspace*{5.0cm}}

\begin{document}

\maketitle %
\pagenumbering{arabic}

\newpage
\phantomsection \addcontentsline{toc}{section}{Contents} %
\tableofcontents

\newpage
\phantomsection \addcontentsline{toc}{section}{List of Figures} %
\listoffigures

%\phantomsection \addcontentsline{toc}{section}{List of Tables} %
%\listoftables

\newpage
\phantomsection \addcontentsline{toc}{section}{Abstract} \noindent
{\noindent \LARGE \bf Abstract} \vspace{0.5cm}\\
\noindent %

Euler-Lagrange variational principle is used to obtain analytical and numerical flow relations in
cylindrical tubes. The method is based on minimizing the total stress in the flow duct using the
fluid constitutive relation between stress and rate of strain. Newtonian and non-Newtonian fluid
models; which include power law, Bingham, Herschel-Bulkley, Carreau and Cross; are used for
demonstration.


%%%%%%%%%%%%%%%%%%%%%%%%%%%%%%%%%%%  Head style  %%%%%%%%%%%%%%%%%%%%%%%%%%%%%%%%%%%
\pagestyle{headings} %
\addtolength{\headheight}{+1.6pt}
\lhead[{Chapter \thechapter \thepage}]%
      {{\bfseries\rightmark}}
\rhead[{\bfseries\leftmark}]%
     {{\bfseries\thepage}} %tell it to put page number at rhead
\headsep = 1.0cm               % Added 07 Sep 2006
%%%%%%%%%%%%%%%%%%%%%%%%%%%%%%%%%%%%%%%%%%%%%%%%%%%%%%%%%%%%%%%%%%%%%%%%%%%%%%%%%%%%

\newpage
%XXXXXXXXXXXXXXXXXXXXXXXXXXXXXXXXXXXXXXXXXXXXXXXXXXXXXXXXXXXXXXXXX
\section{Introduction} \label{Introduction}

Several methods are in use to derive relations between pressure, $p$, and volumetric flow rate,
$Q$, in tubes and conduits. These methods include the application of first principles of fluid
mechanics with utilizing the fluid basic properties \cite{Skellandbook1967, SochiThesis2007,
SochiB2008}, the use of Navier-Stokes equations \cite{SochiNavier2013}, the lubrication
approximation \cite{SochiPower2011}, and Weissenberg-Rabinowitsch-Mooney relation
\cite{Skellandbook1967, CarreaubookKC1997, SochiB2008, SochiThesis2007}. Numerical methods related
to these analytical formulations, such as finite element and similar meshing techniques
\cite{SochiVE2009}, are also in use when analytical expressions are not available.

However, we are not aware of the use of the Euler-Lagrange variational principle to derive $p$-$Q$
relations in general and in capillaries in particular despite the fact that this principle is more
intuitive and natural to use as it is based on a more fundamental physical principle which is
minimizing the total stress combined with the utilization of the fluid constitutive relation
between stress and rate of strain.

The objective of this paper is to outline this method demonstrating its application to Newtonian
and some time-independent non-Newtonian fluids and featuring its applicability numerically as well
as analytically. In the following, we assume a laminar, axi-symmetric, incompressible, steady,
viscous, isothermal, fully-developed flow for generalized Newtonian fluids moving in cylindrical
tubes where no-slip at wall condition \cite{SochiSlip2011} applies and where the flow velocity
profile has a stationary derivative point at the middle of the tube ($r=0$) meaning the profile has
a blunt rounded vertex.

%XXXXXXXXXXXXXXXXXXXXXXXXXXXXXXXXXXXXXXXXXXXXXXXXXXXXXXXXXXXXXXXXX
\section{Method}

The constitutive relation for generalized Newtonian fluids in shear flow is given by

\begin{equation}
\tau=\mu\gamma
\end{equation}
where $\tau$ is the shear stress, $\gamma$ is the rate of shear strain, and $\mu$ is the shear
viscosity which generally is a function of the rate of shear strain. It is physically intuitive
that the flow velocity profile in a tube (or in a flow path in general) will adjust itself to
minimize the total stress which is given by

\begin{equation}
\tau_{t}=\int_{\tau_{c}}^{\tau_{w}}d\tau=\int_{0}^{R}\frac{d\tau}{dr}dr=\int_{0}^{R}\frac{d}{dr}\left(\mu\gamma\right)dr=\int_{0}^{R}\left(\gamma\frac{d\mu}{dr}+\mu\frac{d\gamma}{dr}\right)dr\label{totalStress}
\end{equation}
where $\tau_{t}$ is the total stress, $\tau_{c}$ and $\tau_{w}$ are the shear stress at the tube
center and tube wall respectively, and $R$ is the tube radius. The total stress, as given by
Equation \ref{totalStress}, can be minimized by applying the Euler-Lagrange variational principle
which, in its most famous form, is given by

\begin{equation}
\frac{\partial f}{\partial y}-\frac{d}{dx}\left(\frac{\partial f}{\partial
y'}\right)=0
\end{equation}
where

\begin{equation}
x\equiv r,\,\,\,\,\, y\equiv\gamma,\,\,\,\,\,
f\equiv\gamma\frac{d\mu}{dr}+\mu\frac{d\gamma}{dr},\,\,\,\mathrm{\,\, and}\,\,\,\,\,\frac{\partial
f}{\partial
y'}\equiv\frac{\partial}{\partial\gamma'}\left(\gamma\frac{d\mu}{dr}+\mu\frac{d\gamma}{dr}\right)=\mu
\end{equation}
However, to simplify the derivation we use here another form of the Euler-Lagrange principle which
is given by

\begin{equation}
\frac{d}{dx}\left(f-y'\frac{\partial f}{\partial y'}\right)-\frac{\partial f}{\partial
x}=0
\end{equation}
that is

\begin{equation}
\frac{d}{dr}\left(\gamma\frac{d\mu}{dr}+\mu\frac{d\gamma}{dr}-\mu\frac{d\gamma}{dr}\right)-\frac{\partial}{\partial
r}\left(\gamma\frac{d\mu}{dr}+\mu\frac{d\gamma}{dr}\right)=0
\end{equation}
i.e.

\begin{equation}
\frac{d}{dr}\left(\gamma\frac{d\mu}{dr}\right)-\frac{\partial}{\partial
r}\left(\gamma\frac{d\mu}{dr}+\mu\frac{d\gamma}{dr}\right)=0
\end{equation}

Since ordinary derivative is a special case of partial derivative, we can write this equation as

\begin{equation}
\frac{\partial}{\partial
r}\left(\gamma\frac{d\mu}{dr}-\gamma\frac{d\mu}{dr}-\mu\frac{d\gamma}{dr}\right)=0
\end{equation}
that is

\begin{equation}\label{MainEq}
\frac{\partial}{\partial r}\left(\mu\frac{d\gamma}{dr}\right)=0
\end{equation}

In the following sections the use of this equation will be demonstrated to derive $p$-$Q$ flow
relations for generalized Newtonian fluids.

%XXXXXXXXXXXXXXXXXXXX
\subsection{Newtonian}\label{NewtonianSec}

For Newtonian fluids, the viscosity is constant that is

\begin{equation}
\mu=\mu_{o}
\end{equation}
and hence Equation \ref{MainEq} becomes

\begin{equation}
\frac{\partial}{\partial r}\left(\mu_{o}\frac{d\gamma}{dr}\right)=0
\end{equation}

On integrating once with respect to $r$ we obtain

\begin{equation}
\mu_{o}\frac{d\gamma}{dr}=A
\end{equation}
where $A$ is a constant. Hence

\begin{equation}
\gamma=\frac{1}{\mu_{o}}\left(Ar+B\right)
\end{equation}
where $B$ is another constant. Now from the two boundary conditions at $r=0$ and $r=R$, $A$ and $B$
can be determined, that is

\begin{equation}
\gamma\left(r=0\right)=0\,\,\,\,\,\,\,\,\,\Rightarrow\,\,\,\,\,\,\,\, B=0
\end{equation}
and

\begin{equation}
\gamma\left(r=R\right)=\frac{\tau_w}{\mu_o}=\frac{PR}{2L\mu_{o}}=\frac{AR}{\mu_{o}}\,\,\,\,\,\,\,\,\,\Rightarrow\,\,\,\,\,\,\,\,
A=\frac{P}{2L}\end{equation}
where $\tau_w$ is the shear stress at the tube wall, $P$ is the pressure drop across the tube and
$L$ is the tube length. Hence

\begin{equation}\label{NewtonianGamR}
\gamma\left(r\right)=\frac{P}{2L\mu_{o}}r
\end{equation}

On integrating this with respect to $r$ the standard Hagen-Poiseuille parabolic velocity profile is
obtained, that is

\begin{equation}
v\left(r\right)=\int dv=\int\frac{dv}{dr}dr=\int\gamma
dr=\int\frac{P}{2L\mu_{o}}rdr=\frac{P}{4L\mu_{o}}r^{2}+D
\end{equation}
where $v(r)$ is the fluid axial velocity at $r$ and $D$ is another constant which can be determined
from the no-slip at wall boundary condition, that is

\begin{equation}
v\left(r=R\right)=0\,\,\,\,\,\,\,\,\Rightarrow D=-\frac{PR^{2}}{4L\mu_{o}}
\end{equation}
that is

\begin{equation}
v\left(r\right)=\frac{-P}{4L\mu_{o}}\left(R^{2}-r^{2}\right)
\end{equation}
where the minus sign at the front arises from the fact that the pressure gradient is opposite in
direction to the flow velocity vector. The volumetric flow rate will then follow by integrating the
flow velocity profile with respect to the cross sectional area, that is

\begin{equation}\label{NewtonianQP}
Q=\int_{0}^{R}|v|\,2\pi rdr=\frac{\pi
P}{2L\mu_{o}}\int_{0}^{R}\left(R^{2}-r^{2}\right)rdr=\frac{\pi PR^{4}}{8L\mu_{o}}
\end{equation}
which is the well-known Hagen-Poiseuille flow relation.

%XXXXXXXXXXXXXXXXXXXX
\subsection{Power Law}\label{PowerLawSec}

For power law fluids, the viscosity is given by

\begin{equation}
\mu=k\gamma^{n-1}
\end{equation}

On applying Euler-Lagrange variational principle (Equation \ref{MainEq}) we obtain

\begin{equation}
\frac{\partial}{\partial r}\left(k\gamma^{n-1}\frac{d\gamma}{dr}\right)=0
\end{equation}

On integrating once with respect to $r$ we obtain

\begin{equation}
k\gamma^{n-1}\frac{d\gamma}{dr}=A
\end{equation}

On separating the two variables and integrating both sides we obtain

\begin{equation}
\gamma=\sqrt[n]{\frac{n}{k}\left(Ar+B\right)}
\end{equation}
where $A$ and $B$ are constants which can be determined from the two boundary conditions, that is

\begin{equation}
\gamma\left(r=0\right)=0\,\,\,\,\,\,\,\,\,\Rightarrow\,\,\,\,\,\,\,\, B=0
\end{equation}
and

\begin{equation}
\gamma\left(r=R\right)=\sqrt[n]{\frac{\tau_{w}}{k}}=\sqrt[n]{\frac{PR}{2Lk}}=\sqrt[n]{\frac{n}{k}AR}\,\,\,\,\,\,\,\,\,\Rightarrow\,\,\,\,\,\,\,\,
A=\frac{P}{2nL}
\end{equation}
and therefore

\begin{equation}\label{PowerLawGamR}
\gamma=\sqrt[n]{\frac{P}{2kL}}r^{1/n}
\end{equation}

On integrating this with respect to $r$ the power law velocity profile is obtained, that is

\begin{equation}
v\left(r\right)=\int dv=\int\frac{dv}{dr}dr=\int\gamma
dr=\int\sqrt[n]{\frac{P}{2kL}}r^{1/n}dr=\frac{n}{n+1}\sqrt[n]{\frac{P}{2kL}}r^{1+1/n}+D
\end{equation}
where $D$ is another constant. From the no-slip at wall boundary condition

\begin{equation}
v\left(r=R\right)=0\,\,\,\,\,\,\,\,\Rightarrow D=-\frac{n}{n+1}\sqrt[n]{\frac{P}{2kL}}R^{1+1/n}
\end{equation}
that is

\begin{equation}
v\left(r\right)=\frac{-n}{n+1}\sqrt[n]{\frac{P}{2kL}}\left(R^{1+1/n}-r^{1+1/n}\right)
\end{equation}

The volumetric flow rate will then follow by integrating the velocity profile with respect to the
cross sectional area, that is

\begin{equation}\nonumber
Q=\int_{0}^{R}|v|\,2\pi rdr=\frac{2\pi
n}{n+1}\sqrt[n]{\frac{P}{2kL}}\int_{0}^{R}\left(R^{1+1/n}-r^{1+1/n}\right)rdr
\end{equation}

\begin{equation}\label{PowerLawQP}
=\frac{\pi n}{3n+1}\sqrt[n]{\frac{P}{2kL}}R^{3+1/n}
\end{equation}
which is mathematically equivalent to the expressions derived in \cite{Skellandbook1967,
SochiThesis2007, SochiB2008, SochiPower2011} using other methods.

%XXXXXXXXXXXXXXXXXXXXXXXXXXXXXXXXXXXXXXXXXXXXXXXXXXXXXXXXXXXXXXXXX
\section{Numerical Implementation}

For generalized Newtonian fluids with complex constitutive relations, it may be very difficult, or
even impossible, to obtain a flow analytical solution from the Euler-Lagrange principle. In this
case, the variational method can be used as a basis for a numerical method by employing Equation
\ref{MainEq} to obtain the rate of shear strain as an explicit or implicit function of $r$ which is
then numerically solved and integrated to obtain the flow velocity profile which, in its turn, is
numerically integrated to obtain the $p$-$Q$ relation. For the fluids which have an explicit
relation between the rate of strain and radius, such as Newtonian and power law fluids (refer to
Equations \ref{NewtonianGamR} and \ref{PowerLawGamR}), the rate of strain can be computed directly
for each $r$. However, for the fluids which have no such explicit relation, such as Bingham,
Herschel-Bulkley, Carreau and Cross (refer to Equations \ref{BingGamR}, \ref{HBGamR},
\ref{CarreauGamR} and \ref{CrossGamR}), a numerical solver, based for instance on a bisection
method, is required to find the rate of strain as a function of radius. A numerical integration
scheme; such as midpoint, trapezium or Simpson rule; can then be utilized to integrate the strain
rate with respect to radius to obtain the velocity profile in the first stage, and to integrate the
velocity profile with respect to the cross sectional area to get the volumetric flow rate in the
second stage.

\begin{figure}[!h]
\centering{}
\includegraphics
[scale=0.55] {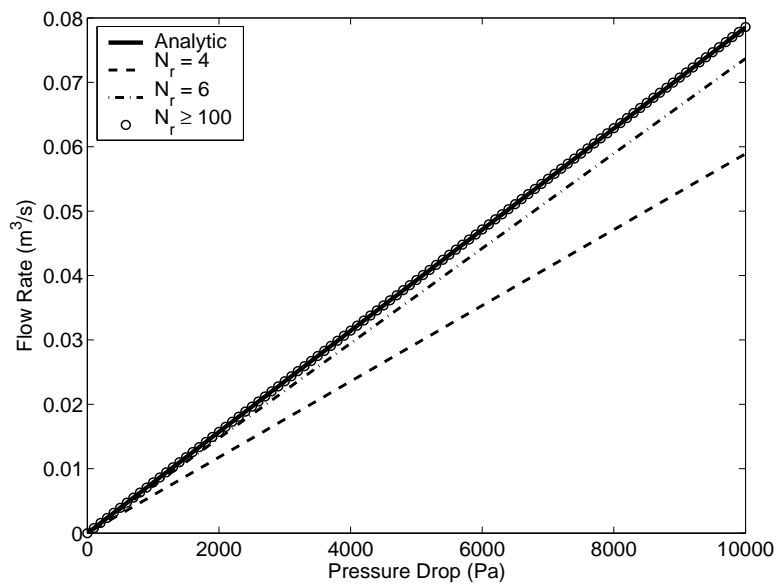} \caption{$Q$ versus $P$ plot for numeric solutions of a typical
Newtonian fluid with $\mu_o=0.005$~Pa.s, flowing in a tube with $L=0.1$~m and $R=0.01$~m for
$r$-discretization $N_r=4$, $N_r=6$ and $N_r\geq100$ alongside the analytical solution (Equation
\ref{NewtonianQP}).} \label{Newtonian}
\end{figure}

\begin{figure}[!h]
\centering{}
\includegraphics
[scale=0.55] {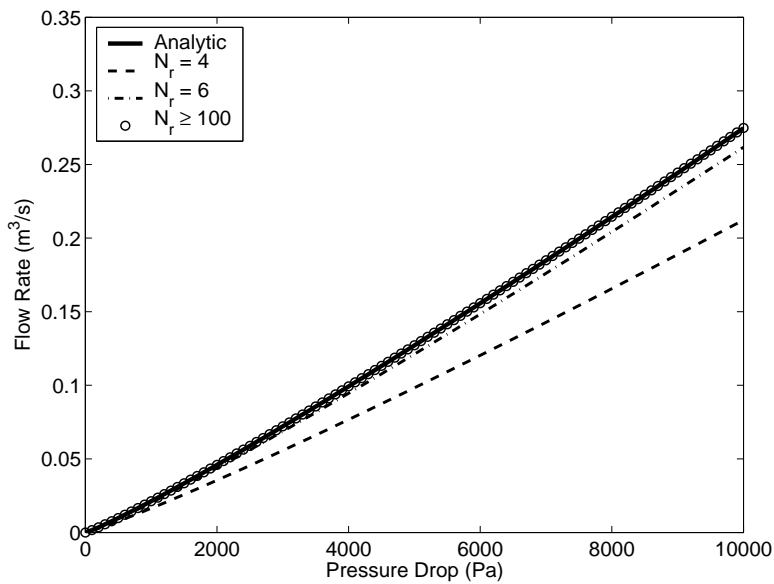} \caption{$Q$ versus $P$ plot for numeric solutions of a typical power law
fluid with $n=0.9$ and $k=0.005$~Pa.s$^n$, flowing in a tube with $L=0.1$~m and $R=0.01$~m for
$r$-discretization $N_r=4$, $N_r=6$ and $N_r\geq100$ alongside the analytical solution (Equation
\ref{PowerLawQP}).} \label{PowerLaw}
\end{figure}

The constant of the first integration (strain rate with respect to radius) is incorporated within a
boundary condition by starting at the wall with zero velocity ($v=0$); the velocity growth at the
next inner ring of the tube cross section, obtained from numerically integrating the shear rate
over radius, is then added incrementally to the velocity of the neighboring previous outer ring to
obtain the velocity at the inner ring. The volumetric flow rate is then computed by multiplying the
velocity of the ring with its cross sectional area and adding these partial flow rate contributions
to obtain the total flow rate. This method is applied, for the purpose of test and validation, to
the Newtonian and power law fluids, for which analytical solutions are available, and the numerical
results were compared to these analytical solutions. A sample of these comparisons are provided in
Figures \ref{Newtonian} and \ref{PowerLaw}. As seen, the numerical solutions converge fairly
quickly to the analytical solutions; hence confirming the reliability of this numerical method and
its theoretical foundations. It should be remarked that the numerical results presented in the
following sections were obtained by using three numerical integration schemes: midpoint, trapezium
and Simpson rules. In all cases the three schemes converged to the same value although with
different convergence rate. For the fluids with an implicit $\gamma$-$r$ relation, a bisection
numerical solver was used to obtain $\gamma$ as a function of $r$.

%XXXXXXXXXXXXXXXXXXXXXXXXXXXXXXXXXXXXXXXXXXXXXXXXXXXXXXXXXXXXXXXXX
\section{Yield Stress Fluids}\label{YieldStressFluids}

Experimenting with the use of Euler-Lagrange principle on different fluids, we tested this method
on some yield stress fluids, specifically Bingham plastic and Herschel-Bulkley models, despite our
awareness of the limitation of this method and its restriction to fluids which makes it
inapplicable to yield stress materials since the solid-like plug flow at the center of the tube
invalidates this assumption.

For Bingham fluids, the viscosity is given by \cite{SochiThesis2007, SochiB2008, SochiComp2010,
SochiYield2010}

\begin{equation}
\mu=\frac{\tau_{o}}{\gamma}+C
\end{equation}
where $\tau_{o}$ is the yield stress and $C$ is the fluid consistency factor. On applying
Euler-Lagrange variational principle (Equation \ref{MainEq}) and following the method outlined in
the Newtonian and power law fluid sections we obtain

\begin{equation}\label{BingGamR}
\tau_{o}\ln\gamma+C\gamma=Ar+B
\end{equation}
where $A$ and $B$ are the constants of integration. As seen in the last equation, the boundary
condition at $r=0$ cannot be used to find $B$ because $\gamma=0$ is a singularity point. We
therefore followed the non-yield stress fluid style and arbitrarily set $B=0$. On applying the
other boundary condition at $r=R$, we obtain

\begin{equation}
A=\frac{\tau_{o}}{R}\ln\left(\frac{PR}{2LC}-\frac{\tau_{o}}{C}\right)+\left(\frac{P}{2L}-\frac{\tau_{o}}{R}\right)
\end{equation}

For Herschel-Bulkley fluids, the viscosity is given by \cite{SochiThesis2007, SochiB2008,
SochiComp2010, SochiYield2010}

\begin{equation}
\mu=\frac{\tau_{o}}{\gamma}+C\gamma^{n-1}
\end{equation}

Following a similar approach to that outlined in the Bingham part, the following relation was
obtained

\begin{equation}\label{HBGamR}
\tau_{o}\ln\gamma+\frac{C}{n}\gamma^{n}=Ar
\end{equation}
where

\begin{equation}
A=\frac{\tau_{o}}{R}\ln\left(\left(\frac{PR}{2LC}-\frac{\tau_{o}}{C}\right)^{1/n}\right)+\frac{1}{n}\left(\frac{P}{2L}-\frac{\tau_{o}}{R}\right)
\end{equation}

\begin{figure}[!h]
\centering{}
\includegraphics
[scale=0.6] {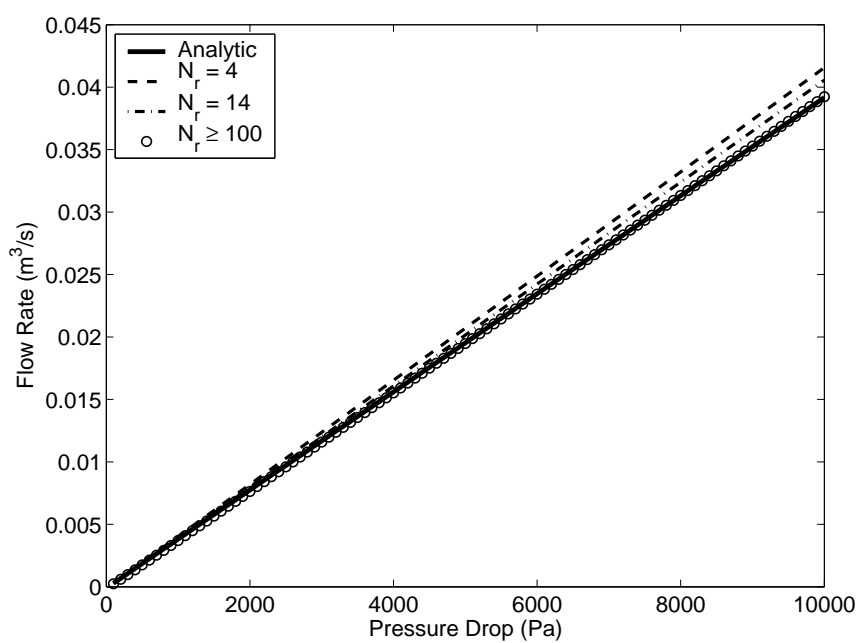} \caption{$Q$ versus $P$ plot for numeric solutions of a typical Bingham
fluid above the yield point with $C=0.01$~Pa.s and $\tau_o=1.0$~Pa flowing in a tube with $L=0.1$~m
and $R=0.01$~m for $r$-discretization $N_r=4$, $N_r=14$ and $N_r\geq100$ alongside the analytic
solution \cite{SochiThesis2007, SochiB2008}.} \label{BinghamLow}
\end{figure}

\begin{figure}[!h]
\centering{}
\includegraphics
[scale=0.6] {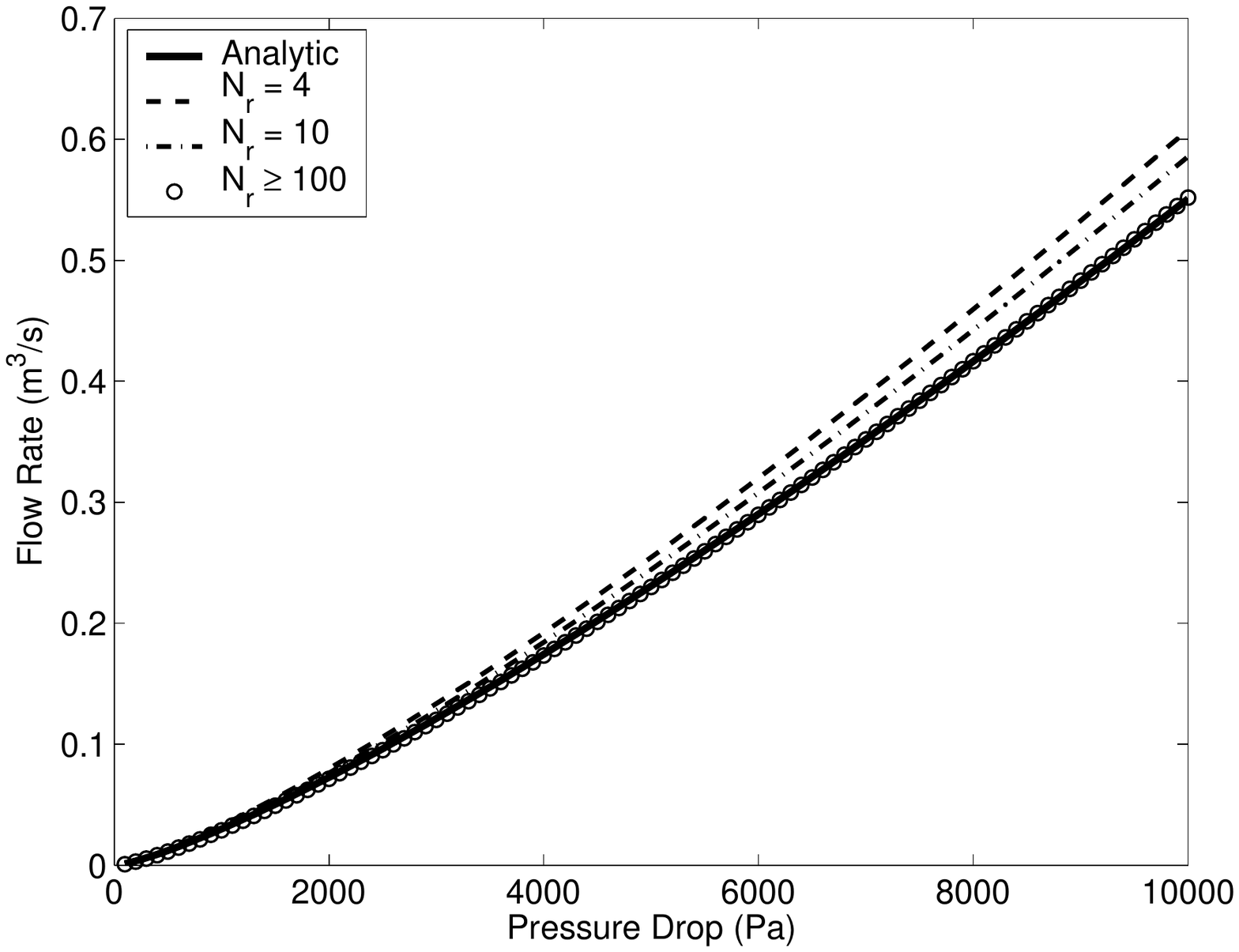} \caption{$Q$ versus $P$ plot for numeric solutions of a typical
Herschel-Bulkley fluid above the yield point with $C=0.01$~Pa.s$^n$, $n=0.8$ and $\tau_o=1.0$~Pa
flowing in a tube with $L=0.1$~m and $R=0.01$~m for $r$-discretization $N_r=4$, $N_r=10$ and
$N_r\geq100$ alongside the analytic solution \cite{SochiThesis2007, SochiB2008}.} \label{HBLow}
\end{figure}

\begin{figure}[!h]
\centering{}
\includegraphics
[scale=0.6] {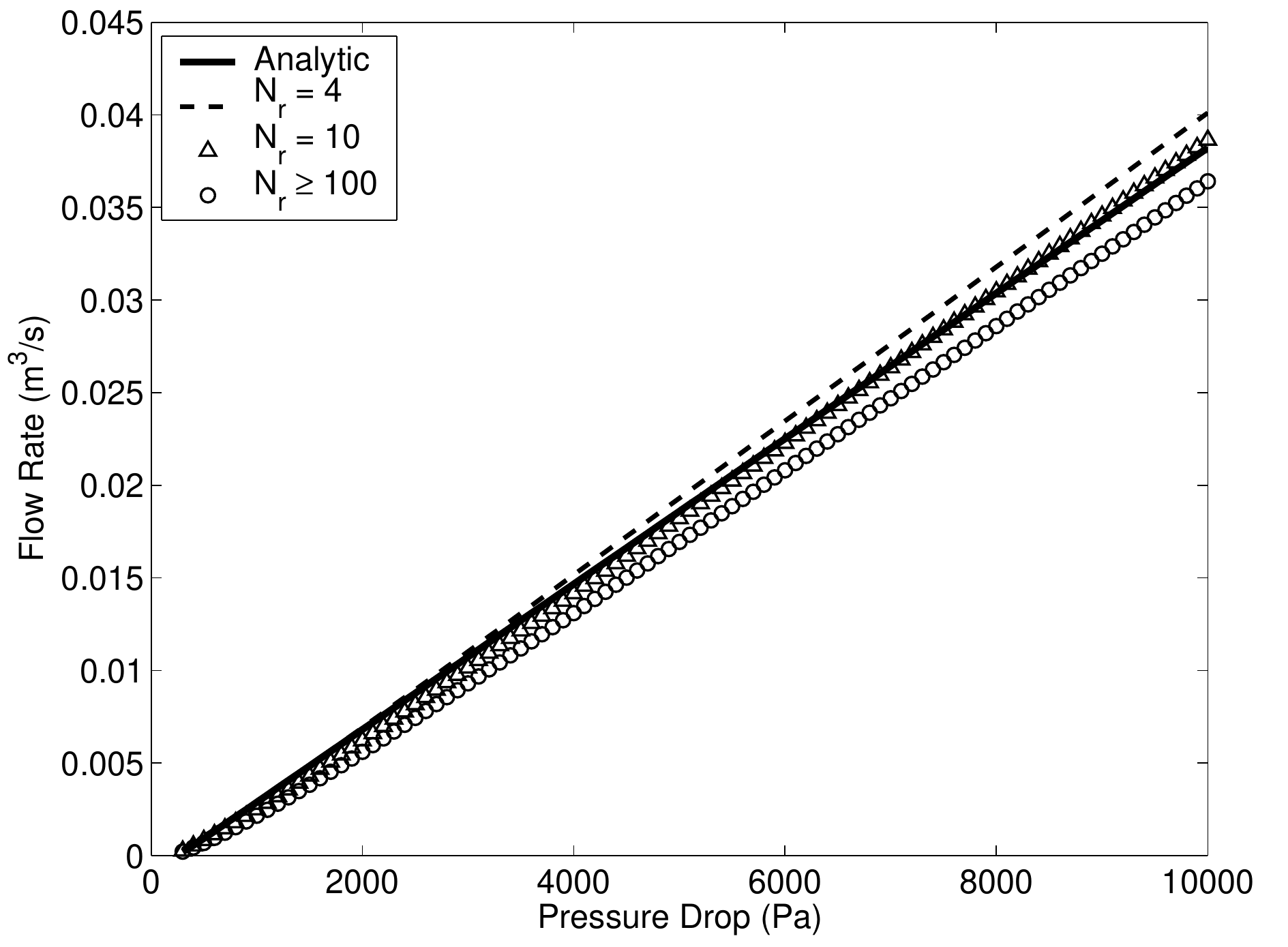} \caption{$Q$ versus $P$ plot for numeric solutions of a typical Bingham
fluid above the yield point with $C=0.01$~Pa.s and $\tau_o=10.0$~Pa flowing in a tube with
$L=0.1$~m and $R=0.01$~m for $r$-discretization $N_r=4$, $N_r=10$ and $N_r\geq100$ alongside the
analytic solution \cite{SochiThesis2007, SochiB2008}.} \label{BinghamHigh}
\end{figure}

\begin{figure}[!h]
\centering{}
\includegraphics
[scale=0.6] {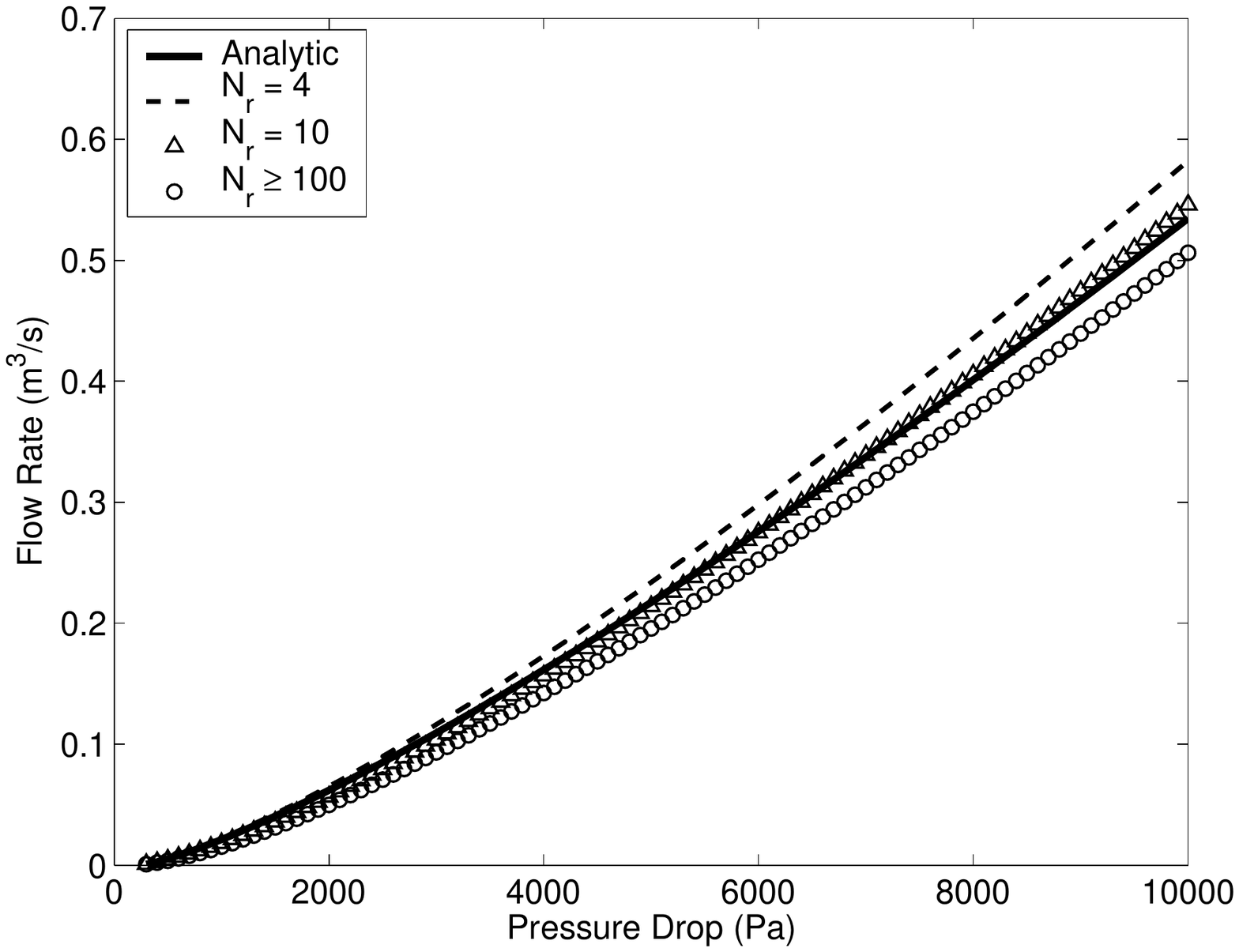} \caption{$Q$ versus $P$ plot for numeric solutions of a typical
Herschel-Bulkley fluid above the yield point with $C=0.01$~Pa.s$^n$, $n=0.8$ and $\tau_o=10.0$~Pa
flowing in a tube with $L=0.1$~m and $R=0.01$~m for $r$-discretization $N_r=4$, $N_r=10$ and
$N_r\geq100$ alongside the analytic solution \cite{SochiThesis2007, SochiB2008}.} \label{HBHigh}
\end{figure}

These $\gamma$-$r$ relations (i.e. Equation \ref{BingGamR} for Bingham and Equation \ref{HBGamR}
for Herschel-Bulkley) were then solved for $\gamma$ at each $r$ and numerically integrated to
obtain the flow velocity profile first and volumetric flow rate second. The numerical results were
interesting as the low yield stress fluids converged correctly to the analytic solution (refer to
Figures \ref{BinghamLow} and \ref{HBLow}) especially at high flow rates, while the high yield
stress fluids diverged with finer discretization (refer to Figures \ref{BinghamHigh} and
\ref{HBHigh}). This can be explained by the occurrence of plug flow at the middle of the tube which
is considerable in the case of high yield stress fluids. The minor deviation of the low yield
stress fluids at low pressures highlights this fact since the plug flow effect diminishes as the
pressure and flow rate increase. The failure of this approach for high yield stress fluids is
simply because our model applies to fluids; and yield stress materials prior to yield at plug area
behave like solids. The convergence of the low yield stress fluids is due to the fact that
negligible plug occurs at the middle of the tube especially at high pressures.

%XXXXXXXXXXXXXXXXXXXXXXXXXXXXXXXXXXXXXXXXXXXXXXXXXXXXXXXXXXXXXXXXX
\section{Carreau and Cross Fluids}\label{OtherFluids}

The Euler-Lagrange variational method was also applied to Carreau and Cross fluids for which no
analytical expressions are available; the details are outlined in the following.

For Carreau fluids, the viscosity is given by \cite{Sorbiebook1991, SochiThesis2007,
SochiArticle2010, SochiFeature2010, Tannerbook2000}

\begin{equation}
\mu=\mu_{\infty}+\left(\mu_{o}-\mu_{\infty}\right)\left[1+\left(\lambda\gamma\right)^{2}\right]^{\left(n-1\right)/2}
\end{equation}
where $\mu_{o}$ is the zero-shear viscosity, $\mu_{\infty}$ is the infinite-shear viscosity,
$\lambda$ is a time constant, and $n$ is the flow behavior index. On applying Euler-Lagrange
variational principle (Equation \ref{MainEq}) and following the derivation, as outlined in \S\
\ref{NewtonianSec} and \ref{PowerLawSec}, we obtain

\begin{equation}\label{CarreauGamR}
\mu_{\infty}\gamma+\left(\mu_{o}-\mu_{\infty}\right)\gamma\,_{2}F_{1}\left(\frac{1}{2},\frac{1-n}{2};\frac{3}{2};-\lambda^{2}\gamma^{2}\right)=Ar+B
\end{equation}
where $_{2}F_{1}$ is the hypergeometric function, and $A$ and $B$ are the constants of integration.
Now from the two boundary conditions at $r=0$ and $r=R$, $A$ and $B$ can be determined, that is

\begin{equation}
\gamma\left(r=0\right)=0\,\,\,\,\,\,\,\,\,\Rightarrow\,\,\,\,\,\,\,\, B=0
\end{equation}
and

\begin{equation}\label{CarBound2}
\gamma\left(r=R\right)=\gamma_{w}\,\,\,\,\,\,\,\,\,\Rightarrow\,\,\,\,\,\,\,\,\mu_{\infty}\gamma_{w}+\left(\mu_{o}-\mu_{\infty}\right)\gamma_{w}\,_{2}F_{1}\left(\frac{1}{2},\frac{1-n}{2};\frac{3}{2};-\lambda^{2}\gamma_{w}^{2}\right)=AR
\end{equation}
where $\gamma_{w}$ is the shear rate at the tube wall. Now, by definition we have

\begin{equation}
\tau_{w}=\mu_{w}\gamma_{w}
\end{equation}
that is

\begin{equation}
\frac{PR}{2L}=\left[\mu_{\infty}+\left(\mu_{o}-\mu_{\infty}\right)\left[1+\left(\lambda\gamma_{w}\right)^{2}\right]^{\left(n-1\right)/2}\right]\gamma_{w}
\end{equation}

From the last equation, $\gamma_{w}$ can be obtained numerically by a numerical solver, based for
example on a bisection method, and hence from Equation \ref{CarBound2} $A$ is obtained

\begin{equation}\label{CarreauA}
A=\frac{\mu_{\infty}\gamma_{w}+\left(\mu_{o}-\mu_{\infty}\right)\gamma_{w}\,_{2}F_{1}\left(\frac{1}{2},\frac{1-n}{2};\frac{3}{2};-\lambda^{2}\gamma_{w}^{2}\right)}{R}
\end{equation}

\begin{figure}[!h]
\centering{}
\includegraphics
[scale=0.55] {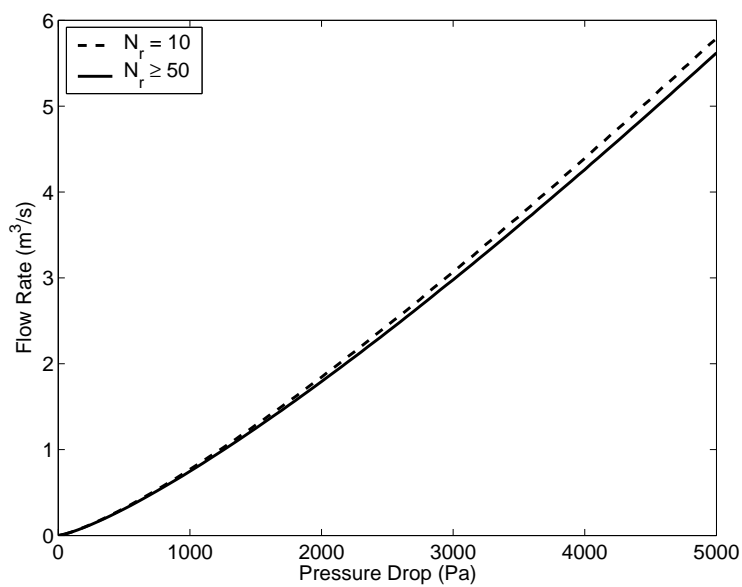} \caption{$Q$ versus $P$ plot for numeric solutions of a typical Carreau
fluid with $n=0.75$, $\mu_o=0.05$~Pa.s, $\mu_{\infty}=0.001$~Pa.s, and $\lambda=1.0$~s flowing in a
tube with $L=0.5$~m and $R=0.05$~m for $r$-discretization $N_r=10$ and $N_r\geq50$. The numeric
solutions were obtained using the real part of the hypergeometric function $_{2}F_{1}$ in Equations
\ref{CarreauGamR} and \ref{CarreauA}.} \label{CarreauQ}
\end{figure}

\begin{figure}[!h]
\centering{}
\includegraphics
[scale=0.55] {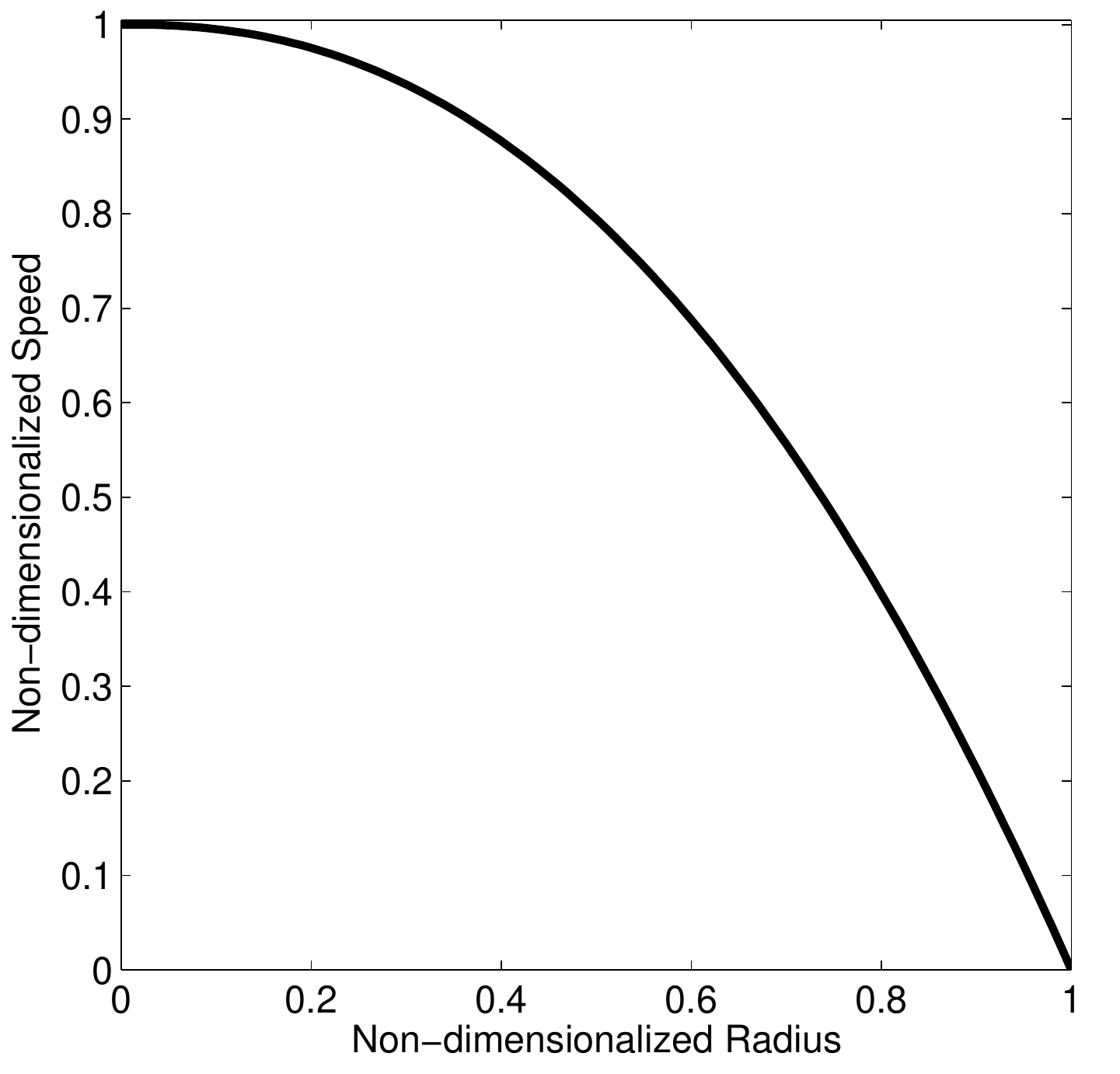} \caption{Flow velocity profile for the Carreau fluid of Figure
\ref{CarreauQ} at a typical flow condition where the radius is scaled to unity and the speed is
scaled to its maximum value at the tube center.} \label{CarreauV}
\end{figure}

Several types of Carreau fluids were used for testing the model and its numerical implementation; a
sample of these tests is presented in Figure \ref{CarreauQ} where the flow did converge for a
radius discretization $N_r\geq50$. A sample of the flow velocity profile across the tube for the
fluid of Figure \ref{CarreauQ} at a typical flow condition is also presented in Figure
\ref{CarreauV}. These figures are qualitatively correct despite the fact that no analytical
solution is available to fully validate the results.

For Cross fluids, the viscosity is given by \cite{OwensbookP2002}

\begin{equation}
\mu=\mu_{\infty}+\frac{\mu_{o}-\mu_{\infty}}{1+\left(\lambda\gamma\right)^{m}}
\end{equation}
where $\mu_{o}$ is the zero-shear viscosity, $\mu_{\infty}$ is the infinite-shear viscosity,
$\lambda$ is a time constant, and $m$ is an indicial parameter. Following a similar derivation
method to that outlined in Carreau, we obtain

\begin{equation}\label{CrossGamR}
\mu_{\infty}\gamma+\left(\mu_{o}-\mu_{\infty}\right)\gamma\,_{2}F_{1}\left(1,\frac{1}{m};\frac{m+1}{m};-\lambda^{m}\gamma^{m}\right)=Ar
\end{equation}
where

\begin{equation}\label{CrossA}
A=\frac{\mu_{\infty}\gamma_{w}+\left(\mu_{o}-\mu_{\infty}\right)\gamma_{w}\,_{2}F_{1}\left(1,\frac{1}{m};\frac{m+1}{m};-\lambda^{m}\gamma_{w}^{m}\right)}{R}
\end{equation}
with $\gamma_{w}$ being obtained numerically as outlined in Carreau.

\begin{figure}[!h]
\centering{}
\includegraphics
[scale=0.55] {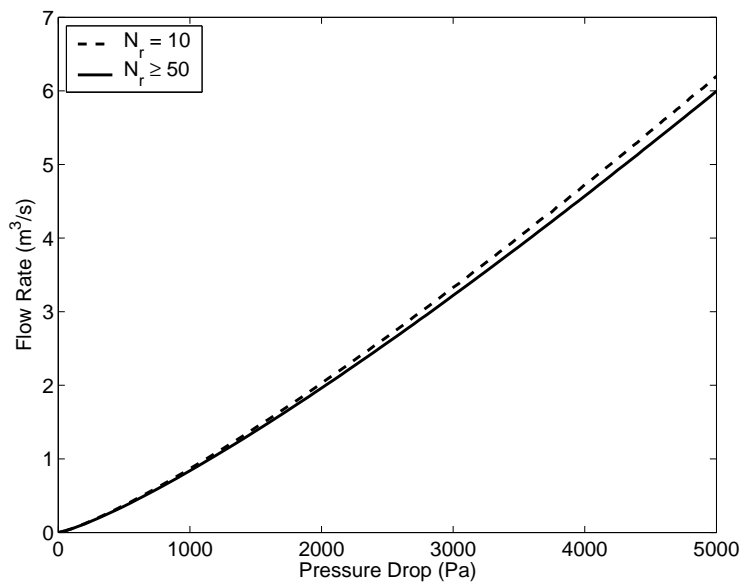} \caption{$Q$ versus $P$ plot for numeric solutions of a typical Cross fluid
with $n=0.25$, $\mu_o=0.05$~Pa.s, $\mu_{\infty}=0.001$~Pa.s, and $\lambda=1.0$~s flowing in a tube
with $L=0.5$~m and $R=0.05$~m for $r$-discretization $N_r=10$ and $N_r\geq50$. The numeric
solutions were obtained using the real part of the hypergeometric function $_{2}F_{1}$ in Equations
\ref{CrossGamR} and \ref{CrossA}.} \label{CrossQ}
\end{figure}

\begin{figure}[!h]
\centering{}
\includegraphics
[scale=0.55] {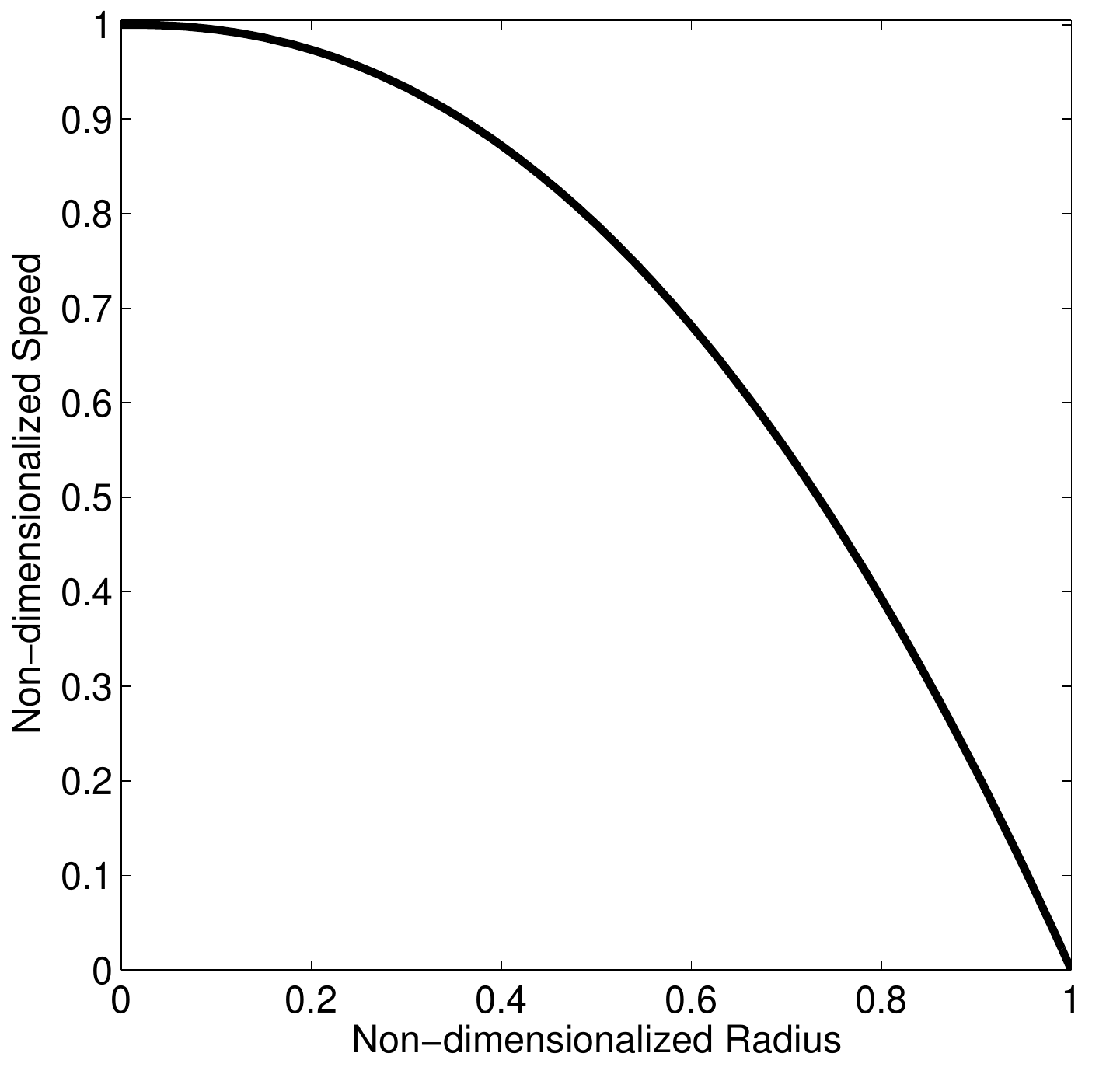} \caption{Flow velocity profile for the Cross fluid of Figure \ref{CrossQ}
at a typical flow condition where the radius is scaled to unity and the speed is scaled to its
maximum value at the tube center.} \label{CrossV}
\end{figure}

Several types of Cross fluids were used for testing the model and its numerical implementation; a
sample of these tests is presented in Figure \ref{CrossQ} where the flow did converge for a radius
discretization $N_r\geq50$. A sample of the flow velocity profile across the tube for the fluid of
Figure \ref{CrossQ} at a typical flow condition is also presented in Figure \ref{CrossV}. These
figures are qualitatively correct despite the fact that no analytical solution is available to
fully validate the results. The features in these results are similar to those observed in Carreau
due to the strong similarities between these two fluids.

\newpage
%XXXXXXXXXXXXXXXXXXXXXXXXXXXXXXXXXXXXXXXXXXXXXXXXXXXXXXXXXXXXXXXXX
\section{Conclusions} \label{Conclusions}

In this article we outlined a method based on Euler-Lagrange variational principle which minimizes
the total stress to obtain analytical and numerical flow relations for generalized Newtonian fluids
in tubes and conduits in general. The method can be used in conjunction with numerical integration
to obtain numerical solutions when analytical integration of the basic equations derived from the
variational principle is difficult or impossible to obtain. The method was validated analytically
and numerically for Newtonian fluids as well as a number of time-independent non-Newtonian fluids.

The main advantages of this method are simplicity, ease of implementation and rapid convergence to
a solution which, for all practical purposes, is identical to the analytical solution. This
convergence can be easily verified from two or more successive $r$-discretization schemes being
converged to the same solution.

The method can be used to obtain flow relations for complex fluids for which no analytical flow
expressions have been derived from other methods due to mathematical difficulties, such as Carreau,
Carreau-Yasuda and Cross. This method, when implemented numerically in the case of analytical
difficulties, is more accurate and suitable than the use of empirical relations or numerical
meshing techniques.

Numerical experiments were performed on Bingham and Herschel-Bulkley yield stress fluids to test
the robustness of this method which is based on the assumption of fluidity. Interestingly, the
method converged correctly to the analytical solution for low yield stress fluids although it did
diverge for high yield stress fluids. The obvious reason which can explain these observations is
that negligible plug flow occurs at the middle of the tube in the first case especially at high
flow rates, while considerable plug flow occurs in the second case which invalidates the basic
assumption of fluidity that this model relies upon.

A preliminary investigation of the applicability of this method to Carreau and Cross fluids has
been conducted and presented in this article. More serious investigations related to these fluids
and other complex fluids by employing this variational technique are planned for the future.

\newpage
%XXXXXXXXXXXXXXXXXXXXXXXXXXXXXXXXXXXXXXXXXXXXXXXXXXXXXXXXXXXXXXXXXXX
\phantomsection \addcontentsline{toc}{section}{Nomenclature} %
{\noindent \LARGE \bf Nomenclature} \vspace{0.5cm}

\begin{supertabular}{ll}
$\gamma$                &   shear rate (s$^{-1}$) \\
$\gamma_w$              &   shear rate at tube wall (s$^{-1}$) \\
$\lambda$               &   time constant (s) \\
$\mu$                   &   fluid dynamic shear viscosity (Pa.s) \\
$\mu_{o}$               &   zero-shear viscosity (Pa.s) \\
$\mu_{\infty}$          &   infinite-shear viscosity (Pa.s) \\
$\tau$                  &   shear stress (Pa) \\
$\tau_c$                &   shear stress at tube center (Pa) \\
$\tau_o$                &   yield stress (Pa) \\
$\tau_t$                &   total shear stress (Pa) \\
$\tau_w$                &   shear stress at tube wall (Pa) \\
\\
$C$                     &   consistency factor in Bingham and Herschel-Bulkley models (Pa.s$^{n}$)                                   \\
$_{2}F_{1}$             &   hypergeometric function \\
$k$                     &   consistency factor in power law model (Pa.s$^n$) \\
$L$                     &   tube length (m) \\
$m$                     &   indicial parameter in Cross model \\
$n$                     &   flow behavior index \\
$N_r$                   &   number of elements in radius discretization \\
$p$                     &   pressure (Pa) \\
$P$                     &   pressure drop (Pa) \\
$Q$                     &   volumetric flow rate (m$^{3}$.s$^{-1}$) \\
$r$                     &   radius (m) \\
%$r_c$                   &   radius at tube center (m) \\
%$r_w$                   &   radius at tube wall (m) \\
$R$                     &   tube radius (m) \\
$v$                     &   axial fluid velocity (m.s$^{-1}$) \\
\end{supertabular}

\newpage
\phantomsection \addcontentsline{toc}{section}{References} %
\bibliographystyle{unsrt}

\end{document}